# OPTIMIZED LAGRANGIAN APPROXIMATIONS
# FOR MODELLING LARGE–SCALE STRUCTURE
# AT NON–LINEAR STAGES

T. Buchert[1], A.L. Melott[2], A.G. Weiss[1]

[1] *Max-Planck-Institut für Astrophysik, Garching, F. R. G.*
[2] *Department of Physics and Astronomy, University of Kansas, Lawrence, U. S. A.*

## ABSTRACT

Approximations to the exact solutions for gravitational instability in the expanding Universe are extremely useful for understanding the evolution of large–scale structure. We report on a series of tests of Newtonian Lagrangian perturbation schemes using N–body simulations for various power–spectra with scale–independent indices in the range $-3$ to $+1$. The models have been evolved deeply into the non–linear regime of structure formation in order to probe the dynamical and statistical performance of the Lagrangian perturbation schemes (whose first–order solution contains as a subset the celebrated "Zel'dovich–approximation" (hereafter ZA). These tests reveal properties of the approximations at stages beyond the obvious validity of perturbation theory. Recently, another series of tests of different analytical and semi–numerical approximations for large–scale structure was conducted with the result that ZA displays the best dynamical performance in comparison with the N–body simulations, if the initial data were smoothed before evolving the model, i.e., a truncated form of ZA (TZA). We show in this contribution that the excellent performance of TZA can be further improved by going to second order in the Lagrangian perturbation approach. The truncated second–order Lagrangian scheme provides a useful improvement over TZA especially for negative power indices, which suggests it will be very useful for modelling standard scenarios such as "Cold–", "Hot–" and "Mixed–Dark–Matter".

## 1. Lagrangian perturbation theory put into perspective

Zel'dovich[1,2] proposed an approximation (hereafter ZA) by extrapolating the Eulerian linear theory of gravitational instability into the non–linear regime using the Lagrangian picture of continuum mechanics. He discussed interesting consequences of this approximation which is capable of describing shell–crossing singularities which, in this model, develop into highly anisotropic oblate "pancake" structures. Those structures can be fundamentally understood and classified in the framework of the Lagrange–singularity–theory, (Arnol'd[3] *et al.*, Shandarin and Zel'dovich[4], Buchert[5] *et al.* and references therein). Zel'dovich's work initiated many applications making it one of the most cited articles in astronomy: it emerged as a standard tool to model principal elements of the large–scale structure, is used to initialize most N–body codes employed by the cosmology community, and forms the basis of more sophisticated approximations like the adhesion approximation (Gurbatov[6] *et al.* , Kofman[7] *et al.* ). Zel'dovich's model can be derived by formulating the Euler–Poisson system in terms of Lagrangian coordinates and solving the Lagrangian evolution equations for the field of trajectories perturbatively (Buchert and Götz[8], Buchert[9,10]). It appears as a subclass of the irrotational Lagrangian first–order solution which covers substantial non–linearities in contrast to the Eulerian first–order solution. This explains the success of this approximation if applied to non–linear gravitational structure formation (compare Coles[11] et al.).

The particular justification that ZA could be relevant to hierarchical clustering has developed slowly (Melott[12] et al. ; Melott and Shandarin[13]; Kofman[14]; Little[15] et al. ; Kofman[7] et al. ; Coles[11] et al. ; however see Peebles[16]). A general concensus is forming based around a unification of the former Soviet ("pancake") and Western ("hierarchical clustering") theories.

The Lagrangian theory of gravitational instability is now used in large–scale structure modelling as much as the Eulerian theory of gravitational instability used to be. Numerous efforts concern the investigation and application of Lagrangian perturbation solutions up to the third order (Buchert[9,10], Moutarde[17] et al. , Bouchet[18] et al. , Buchert[19], Buchert and Ehlers[20], Gramann[21], Giavalisco[22] et al. , Lachièze–Rey[23,24], Buchert[25], Bernardeau[26], Munshi and Starobinsky[27], Munshi[28] et al. , Bouchet[29] et al. ), and, most recently, the investigation of general relativistic analogues (which are intrinsically Lagrangian in the eigensystem of the flow).

## 2. Optimization of Lagrangian perturbation schemes

Until recently, ZA was evolved only until shell–crossing, i.e., when singularities in the density field develop, at the epoch when the Eulerian representation of the basic dynamical equations breaks down. In principle, the Lagrangian representation of the flow allows following the evolution across caustics, where the flow field itself remains finite. This implies neglecting self–gravitating interaction of multi–stream systems developing inside caustics; however, secondary generations of shell–crosssings can be modelled as observed in N–body simulations by going to higher orders in the perturbation approach (Buchert and Ehlers[20]). Melott[30] et al. (hereafter MPS) investigated the performance of a new approximation which requires truncation of high frequencies in the initial power–spectrum before evolving ZA (hereafter: TZA) taking the evolution of large–scale structure deeply into the non–linear regime. MPS found that filtering the initial data with a Gaussian at a scale close to but smaller than the non–linearity scale yields the best agreement with the density fields of the same (untruncated) initial data as evolved by an N–body code.

The non–linearity scale $k_{nl}$ is defined by:

$$a^2(t) \int_0^{k_{nl}(t)} d^3k \, \mathcal{P}(k) \; = \; 1 \; , \qquad (1)$$

where $k_{nl}(t)$ is decreasing with time as successively larger scales enter the non–linear regime; $a(t)$ is the scale factor of the homogeneous background ($a(t_i) \equiv 1$), and $\mathcal{P}(k)$ denotes the initial power–spectrum taken to be a powerlaw with indices in the range $-3$ to $+1$.

"Best agreement" was defined in terms of an optimal scale $k_{opt}$ in $k$–space at which the usual cross–correlation coefficient $S$ between the resulting density fields attains

its maximum:
$$S := \frac{<(\delta_1 \delta_2)>}{\sigma_1 \sigma_2} \quad , \tag{2}$$

where $\delta_\ell, \ell = 1, 2$ represent the density contrasts in the analytical and the numerical approximations, respectively, $\sigma_\ell = \sqrt{<\delta_\ell^2> - <\delta_\ell>^2}$ is the standard deviation in a Gaussian random field; averages $< ... >$ are taken over the entire distribution. We believe this is the most important statistical test, because it measures whether the approximation is moving mass to the right place, with an emphasis on dense regions. We also allow for small errors by calculating $S$ for the two density arrays smoothed at a variety of smoothing lengths.

For the Lagrangian perturbation schemes up to the third order which were used in our tests see Buchert[25]. We conducted several tests: In the first step we studied "pancake models", i.e., models which a priori have a truncated power–spectrum, in order to study principal effects of a second– and higher–order correction to ZA (for details see Buchert[31] et al. ). In the second step we analyzed the whole family of models with powerlaw–spectra $-3, \ldots, +1$ by evolving them deeply into the non–linear regime (for details see Melott[32] et al.). These "hierarchical models" have been evolved for expansion factors of 240 to 5100, depending on spectral index and $k_{nl}$.

Besides the cross–correlation coefficient (2) as a function of scale, we analyzed several statistics including the comparison of the evolved power–spectrum, the evolved r.m.s. values of the density contrast as a function of scale, the phase–angle accuracy achieved by the analytical models, and the evolved density distribution functions.

For all these statistics and for all spectra studied with different filter types and filter scales, we always found improvement for the second–order scheme upon first–order (TZA), if the initial data are truncated with a *Gaussian filter* at a slightly larger scale $k_{opt}$ than the scale needed for the optimal TZA.

For illustrations we refer to Melott[32] et al., where the results of the cross-correlation statistics as well as other statistics are presented in great detail. Especially we want to point at Table 1 and Figure 6 of Melott[32] et al., which present the optimal truncation scales for the various models and the cross correlations between optimized Lagrangian perturbation schemes and numerical simulations, respectively.

## 3. Conclusions

We summarize our main conclusions and list the advantages of going to second–order in perturbation theory for the purpose of modelling highly non–linear stages:

1. The statistics which probe the gravitational dynamics of the models show improvement due to second–order corrections. This success is found for a considerably higher non–linearity than is expected from a perturbation approach.

2. The improvement (although minor for much small–scale power) is *robust* by going to later stages and to smaller scales. This holds for any spectrum and for any statistics analyzed.

3. The CPU times on a CRAY YMP are for the first–order scheme 25 seconds, and for the second–order scheme 60 seconds; the corresponding CPU times on a CONVEX C220 are 2 and 5 minutes. Thus, even the second–order scheme is competitive with *one step* in a corresponding PM–type N–body simulation.

4. The high speed as well as the fact that the second–order scheme is as easy to implement as the first–order scheme (directly from the initial data), render this model suitable for all areas of application where thus far ZA was used, e.g., the initialization of N–body codes.

5. The second–order scheme predicts much faster collapse of first objects (treating also tidal effects) at times comparable to the collapse time in the widely used spherical "tophat" model (Moutarde[17] *et al.*, Buchert and Ehlers[20], Munshi[28] *et al.*, Buchert[34] *et al.*). Thus, it is preferred for the treatment of ensembles of collapsing objects and for normalization purposes.

6. Since the second–order corrections to TZA provide noticeable improvement of dynamical accuracy for initial data with negative sloped power–spectra, we expect that the truncated second–order scheme will be especially useful for the modelling of standard cosmogonies (like "Hot"–, "Cold"–, and "Mixed–Dark–Matter").

7. This modelling will be effective for large sample calculations, since in numerical realizations of 'fair' samples in excess of $300h^{-1}$ Mpc, performed with the same resolution as the simulations reported here ($128^3$ particles on $128^3$ meshs), the truncation scale is close to the Nyquist frequency of the N–body computing. Thus, shortcomings of the analytical schemes become negligible which puts them in an ideal position for the purpose of simulating the environment of galaxy formation down to scales where other physical effects start to affect models based on the description of self–gravity alone. Our method can be effective down to galaxy group mass scales ($10^{13} M_\odot$), or better if we include biasing or go to epochs earlier than the present. Formerly, such approximations have only been used down to about $10^{15} M_\odot$ (see, e.g., Dekel[33]). Thus many things which formerly were studied by N–body methods can now be studied by approximation.

8. The third–order scheme does not show the 'robustness' observed for second– and first–order. However, to draw definite conclusions the *analytical* solution of the third–order effect must be studied with reduced numerical uncertainties in its realization by Fast–Fourier–Transform, as pursued by Buchert[34] *et al.*

The code for second–order is available on request from tob @ mpa-garching.mpg.de

## 4. Acknowledgements

TB was supported by DFG (Deutsche Forschungsgemeinschaft). ALM wishes to acknowledge support from NASA NAGW–3832 and from NSF grants AST–9021414 and OSR–9255223, and facilities of the National Center for Supercomputing Applications, all in the USA. We are grateful to the Aspen Center for Physics (USA) for its June 1994 workshop on topics related to this work.